\documentclass[useAMS,usenatbib]{mn2e}
\usepackage[]{txfonts}
\usepackage{soul}
\usepackage{blindtext}
\usepackage{threeparttable}
\def\simless{\mathbin{\lower 3pt\hbox
{$\rlap{\raise 5pt\hbox{$\char'074$}}\mathchar"7218$}}}   
\def\simmore{\mathbin{\lower 3pt\hbox
{$\rlap{\raise 5pt\hbox{$\char'076$}}\mathchar"7218$}}}   
\newcommand{\be}{\begin{equation}}
\newcommand{\ee}{\end{equation}}
\newcommand{\eqb}{\begin{eqnarray}}
\newcommand{\eqe}{\end{eqnarray}}
\newcommand{\unit}[1]{\nobreak{\mathrm{\;#1}}}

\newcommand{\sth}{\sigma_{\rm T}}

\newcommand{\nup}{$\nu_{\rm p}$}

\newcommand{\src}{Ap~Librae\xspace}

\newcommand{\swift}{{\it Swift}\xspace}

\newcommand{\fermi}{{\it Fermi}-LAT \xspace}
\newcommand{\hess}{H.E.S.S. \xspace}

\newcommand{\hcm}[1]{$\times 10^{#1}$ cm$^{-2}$}

\newcommand{\ergcm}[1]{$\times 10^{#1}$ erg cm$^{-2}$ s$^{-1}$}
\usepackage{graphicx}
\usepackage{epsfig} 
\usepackage{epstopdf}
\usepackage{longtable}
\usepackage{amssymb}
\usepackage{xcolor}
\usepackage{color}
\usepackage{lipsum}
\usepackage{textcomp}
\usepackage{threeparttable}
\usepackage[colorlinks=true,linkcolor=blue,citecolor=blue,
bookmarks=true,bookmarksopen=true,bookmarksnumbered=true,draft=false]{hyperref}
\usepackage{enumerate}

\title[The $\gamma$-ray emission of \src]
{The TeV emission of \src: a hadronic interpretation and prospects for CTA}

\author[Petropoulou, Vasilopoulos, Giannios]
{M. Petropoulou$^{1}$\thanks{E-mail: mpetropo@purdue.edu}\thanks{Einstein Fellow}, G. Vasilopoulos$^2$\thanks{E-mail: gevas@mpe.mpg.de} \& D. Giannios$^{1}$\\
$^{1}$Department of Physics and Astronomy, Purdue University, 525 Northwestern
Avenue, West Lafayette, IN 47907, USA\\
$^{2}$Max-Planck-Institut f\"ur extraterrestrische Physik,Giessenbachstra{\ss}e, 85748 Garching, Germany}

\begin{document}
\date{Received / Accepted}
\pagerange{\pageref{firstpage}--\pageref{lastpage}} \pubyear{2013}

\maketitle

\label{firstpage}

\begin{abstract}
Ap Librae is one out of a handful of low-frequency peaked blazars to be detected at TeV $\gamma$-rays {and the only one with an identified X-ray jet.}
Combined observations of \fermi at high energies (HE)
and of \hess at very high energies (VHE)
revealed a striking spectral property of \src; the presence of a {broad} high-energy component that extends more than nine orders of magnitude in energy and is, therefore, hard to be explained by the usual single-zone synchrotron self-Compton model.
We show that the superposition of different emission components related to photohadronic interactions can explain the $\gamma$-ray emission  of \src without invoking external radiation fields. We present two indicative model fits to the spectral energy distribution of \src where the VHE emission is assumed to originate from a compact, sub-pc scale region of the jet. A robust prediction of our model is VHE flux variability on timescales similar to those observed at X-rays and HE $\gamma$-rays, which can be further used to distinguish between a sub-pc or kpc scale origin of the TeV emission. {We thus} calculate the expected variability signatures at  X-rays, HE and VHE $\gamma$-rays and show that quasi-simultaneous  flares are expected, with larger amplitude flares appearing at $\gamma$-rays.  We assess the detectability
of VHE variability from \src with CTA, {next generation of IACTs.} We show that $\sim$hr timescale variability at $E_{\gamma}>0.1$~TeV could be detectable at high significance with shorter exposure times than current Cherenkov telescopes.
\end{abstract} 
  

\section{Introduction} 
\label{sec:intro}
Blazars with extremely weak optical emission lines or, in many cases,
featureless optical spectra, are classified as BL Lac { objects}. 
The majority of BL Lac {objects} that are detected at very high energies (VHE, $E_\gamma > 100$~GeV)  by ground-based Cherenkov telescopes belongs to 
the high-frequency peaked (HBL) subclass {\citep[e.g.][]{tevcat_08, hinton_09}} that is characterized by a low-energy spectral component peaking at frequencies \nup$> 10^{15}$~Hz \citep{padovani_giommi95}. 
The quiescent emission as well as  individual flares from TeV-detected blazars have been
successfully explained by the synchrotron self-Compton (SSC) model \citep[e.g.][]{maraschietal92, bloommarscher96, mastkirk97, konopelko03, celotti_ghisellini08, weidinger_spanier10, hess_1es1312423_2013, boettcher_reimer13}. In this scenario, electron synchrotron radiation is invoked to explain the low-energy hump of the {spectral energy distribution (SED)}, while it serves as the seed photon field for inverse Compton (IC) scattering, which, in turn, results in the observed high-energy emission. Despite its success the SSC model faces difficulties in explaining the observed emission from certain low-frequency peaked (LBL; \nup$<10^{14}$~Hz) and intermediate-frequency peaked (IBL; $10^{14}$~Hz$<$\nup$< 10^{15}$~Hz)  sources. {Examples of VHE emitting LBL that challenge the SSC interpretation are BL Lacartae \citep{ravasio02} and W Comae \citep{acciari09} (see also \cite{boettcher_reimer13})}. One of the most representative sources {though} is the LBL \src (alternative name {QSO\,B1514-24}).

\src, lying at a redshift $z=0.049\pm0.002$ \citep{disney74,jones09}, belongs to a handful of LBL that have been detected at VHE, such as BL Lacartae \citep{albert07} and S5~0716+714 \citep{anderhub09}, {while it is the only TeV BL Lac object with a detected X-ray jet \citep{kaufmann11, kaufmann13}}\footnote{{For a review on the extended kpc-scale jets of radio-loud AGN, see \cite{harris06}.}}. The first detection of \src at $E_{\gamma}>100$~GeV  by \hess was reported in 2010 \citep{hofmann10} while the resulting observations have been  recently presented in \cite{abramowski15}. These, in combination with the {X-ray and} high-energy (HE) \fermi observations (100 MeV--300 GeV), revealed an unexpected broad high-energy hump, spanning more than nine orders of magnitude in energy {(from X-rays to TeV $\gamma$-rays)}. As first noted by \citet{fortin10}, \citet{tavecchio10} and later by \citet{sanchez12}, the SED of \src cannot be explained by the usual SSC model due to the extreme broadness of the 
high-energy component.

As the SSC model falls short in explaining the broadband spectrum of \src, alternative scenarios have been proposed \citep{hervet_boisson15, sanchez15, zacharias16}. In general terms, these models invoke IC scattering of external photons fields (EC) in order to fill in the gap between the hard X-rays and VHE $\gamma$-rays, caused by the SSC cutoff at energies below the \fermi band.  In particular, \citet{hervet_boisson15} showed that 
photons from the {Broad Line Region (BLR)} may be up-scattered by electrons in the jet at sub-pc scales (blob) making a significant contribution to the \fermi band, while the external IC emission from the interaction of the blob electrons with the pc-jet radiation contributes the most at the \hess energy band. The SED of \src was successfully described from GHz frequencies up to VHE $\gamma$-rays at the cost, however, of increased complexity and number of free parameters.

A second scenario invokes the kpc-scale jet of \src
that has been observed both in radio and X-rays \citep{kaufmann13}. In this case, the VHE emission is explained by IC scattering of the cosmic microwave background (CMB) radiation by relativistic electrons of the kpc jet \citep{sanchez15, zacharias16}. A robust prediction of this model is therefore the lack of variability at VHE in \src. In this scenario the jet is assumed to remain highly relativistic, i.e. with bulk Lorentz factors $\Gamma \gg 1$, even at kpc scales. It is noteworthy that \fermi observations have recently ruled out the IC/CMB interpretation of the X-ray emission from the kpc jets of other powerful AGN \citep{meyer14,meyer15}.

Recently, \citet{petromast15} (henceforth, PM15) showed that broad HE spectra with significant curvature can be obtained outside the usual SSC and EC framework, if protons are accelerated at moderate energies $\sim 10^{16}-10^{17}$~eV, i.e. just above the ``knee'' of the cosmic-ray spectrum (for possible particle acceleration mechanisms in blazars, see e.g. \citealt{biermannstrittmatter87, giannios10, sironi13,sironi15}).
The broadness of the spectra produced in this scenario can be understood as follows. In BL~Lacs, the most important target photon
field for photohadronic ($p\gamma$) interactions with the relativistic protons is the synchrotron radiation of the co-accelerated (primary) electrons, which emerges as the low-energy hump of their SED. The $p\gamma$ interactions are comprised of two processes of astrophysical interest, namely the Bethe-Heitler pair production ($pe$) and  the photopion production  ($p\pi$). Moreover, high-energy photons produced by the decay of neutral pions can be absorbed in the source, thus giving rise to an electromagnetic cascade which peaks at lower energies \citep[][]{mannheim91, mannheim93}. Both processes increase the relativistic pair content of the emission region by injecting {electron-positron} pairs  ($e^{-}e^{+}$), which, in turn, undergo synchrotron and IC cooling as primary electrons. Their radiative signatures will be, in principle, imprinted on the blazar's SED  \citep[see also][]{petrodimi15}.

PM15 showed, in particular, that the synchrotron emission from $pe$ pairs may have important implications for blazar emission.
The broad injection energy distribution of $pe$ pairs \citep[e.g.][]{kelneraharonian08} is also reflected at their synchrotron spectrum, which also appears extended and curved. The $pe$ synchrotron spectrum might appear as a broad component in the range of hard X-rays ($\gtrsim 40$~keV) to soft $\gamma$-rays ($\lesssim 400$~MeV) for certain parameter values (see e.g. Figs.~ 4 and 5 in PM15). Given that the secondary pairs from $p\pi$ interactions are injected with a different rate and at different energies than the Bethe-Heitler secondaries \citep[][]{kelneraharonian08, DMPR2012}, a broad range of HE spectra is expected \citep[e.g.][]{petrodimi15, cerruti15}. 

In this paper, we show that the superposition of  different emission components related to $p\gamma$ interactions can explain the HE and VHE $\gamma$-ray emission  of \src without invoking external radiation fields. Most important is, though, that VHE variability is a robust prediction of our model that can be used to distinguish between a sub-pc or kpc-scale origin of the VHE emission in \src. Being an important diagnostic tool, we calculate first the expected variability signatures at different energy bands (X-rays, HE and VHE $\gamma$-rays) and then focus on VHE $\gamma$ rays.
In this paper we assess the detection of VHE variability from \src with the future TeV Cherenkov Telescope Array (CTA; \citealt{actis_11}) which is designed to surpass the sensitivity of current TeV telescopes. 

This paper is structured as follows.  In Sec.~\ref{sec:data} we present the multi-wavelength data used for the SED compilation and in Sec.~\ref{sec:model} we outline the adopted model.  The results of our model application to \src are presented in Sec.~\ref{sec:results}. In the same Section we present the model predictions on the multi-wavelength variability, while focusing on the VHE $\gamma$-rays. The prospects for CTA are presented in Sec.~\ref{sec:cta}. We continue in Sec.~\ref{sec:discussion} with a discussion of our results and conclude in Sec.~\ref{sec:summary}  with a summary. 
For the required transformations between the reference systems of the blazar and the observer, 
we have adopted a cosmology with $\Omega_{\rm m}=0.3$, $\Omega_{\Lambda}=0.7$ and $H_0=70$ km s$^{-1}$ Mpc$^{-1}$. The redshift of \src \ $z=0.049$  corresponds to a luminosity distance $D_{\rm L}=217.7$~Mpc. The extragalactic background light (EBL) model of \citet{franceschini08} was used for the attenuation of the model-derived $\gamma$-ray spectra. 

\section{Data}
\label{sec:data}
The high-energy spectrum of \src composed by \fermi and \hess observations has recently been presented by \citet{abramowski15}.
The authors analyzed five years of \fermi data (MJD 54682-56508) and divided
their analysis into a quiescent period (MJD 54682-56305 and MJD  56377-56508) and a flaring period (MJD 56306-56376). 
The VHE observations by \hess were performed during the period MJD 55326-55689, i.e. within the quiescent period of \fermi observations. 
No \hess data have been presented for the period of the \fermi flare and no results regarding the variability at VHE have been published by the time of writing. We complement our analysis and SED modelling with multi-wavelength observations that were also temporally coincident with the  \fermi  quiescent period. Data from the flaring period of \src in GeV energies are therefore exlcuded from the present analysis.

We searched the {\tt HEASARC} archive\footnote{\url{http://heasarc.gsfc.nasa.gov/docs/archive.html}} for available X-ray observations of \src \, {during the quiescent period of \fermi observations (MJD 54682-56305).}
The source was observed seven times by \swift (obs-ids: 00036341005-11, MJD 55247-55783) and four times by {\tt RXTE} (MJD 55387.9-55391.8) during the quiescent period defined above. \swift/XRT products were downloaded from  the {\tt HEASARC} archive and analysed following the standard procedures described in the \swift data analysis guide\footnote{\url{http://www.swift.ac.uk/analysis/xrt/}}. 
{\swift/XRT products were generated with the {\tt xrtpipeline} and the events were extracted using the command line interface {\tt xselect}. The auxiliary response files were produced with {\tt xrtmkarf}. For the \swift/XRT analysis the latest response matrix was used, as provided by the \swift calibration database {\tt caldb}.} {\tt RXTE} publication-grade data products were downloaded by the {\tt HEAVENS} webpage\footnote{\url{http://www.isdc.unige.ch/heavens/}}. The X-ray spectra were analysed with {\tt xspec} \citep[version 12.8.2, ][]{1996ASPC..101...17A}. The spectra were fitted simultaneously with an absorbed power-law  model with the addition of a scaling factor to account for variability and instrumental differences.
The X-ray absorption was modelled using the {\tt tbnew} code, a new and improved version of the X-ray absorption model {\tt tbabs} \citep{2000ApJ...542..914W}, while the atomic cross sections were adopted from \citet{1996ApJ...465..487V}. The column density was fixed based on the value provided by the Leiden/Argentine/Bonn (LAB) Survey of Galactic HI \citep[N$_{\rm H{\rm, GAL}}$ = 0.81\hcm{21},][]{2005A&A...440..775K}.  \swift/XRT spectral fitting showed evidence of small flux variability ($<20\%$)  {but no signs of spectral variability, with the average photon index of the absorbed power-law fit being $\Gamma=1.61\pm0.09$. All the above are consistent with the findings of \citet{kaufmann13}.} For the SED presentation, we plot the \swift/XRT spectra with the minimum {(4.6\ergcm{-12})} and maximum {(5.4\ergcm{-12})} derived fluxes from all the available \swift/XRT observations.

\src was observed by the \swift ultraviolet and optical telescope (UVOT) with all its available filters. 
We analysed the \swift/UVOT images and derived the corresponding magnitudes of \src. 
Additional optical observations of the system were performed by the FERMI/SMARTS project \citep{2012ApJ...756...13B}. We used the averaged B, R, J and K magnitudes provided by the SMARTS database\footnote{\url{http://www.astro.yale.edu/smarts/glast/home.php}} for the quiescent period of \src.
All the observed fluxes were corrected for extinction following \citet{1989ApJ...345..245C} where the reddening was set to $E(B-V) = 0.1177$ \citep{2011ApJ...737..103S}. For consistency, we cross-checked our results for the \swift/XRT, UVOT and SMARTS fluxes with the published values reported by \citet{sanchez15} and found that they are compatible.

\src has been monitored at 15 GHz by the Very Long Baseline Array (VLBA) as part of the MOJAVE program\footnote{\url{http://www.physics.purdue.edu/MOJAVE/}}. The radio emission originates from the pc-scale jet and shows no signs of variability within the period MJD 53853-55718 covered by the MOJAVE observations \citep{sanchez15}. In our analysis we adopt the time-averaged flux at 15 GHz as obtained by \citet{sanchez15}. At higher radio frequencies (30 -- 353 GHz) we used the Planck measurements from the Early Release Compact Source Catalog (ERCSC, \citet{PlanckVII_11}) and we included the data at 3.4, 4.6, 12, and 22 $\mu$m from the Wide-field Infrared Survey Explorer (WISE, \citet{WISE_10}). 

The compilation of the multi-wavelength SED of \src was completed with the inclusion of archival data from the NED database \citep{Dixon_70, Kuehr_81, Wright_90, Wright_94, VLA_98,ROSAT_99,CRATES_07,Cusumano_10,Cusumano_10b,Murphy_10,GALEX_11,Giommi_12,1SWXRT_13,Planck_13, 1SXPS_14,ROSAT2_16}.
\section{model}
\label{sec:model}
{In this paper we focus on the emission from the core region of \src. The contribution of the extended jet to the total X-ray flux is $\lesssim 10\%$ \citep{kaufmann13, sanchez15} and can be, therefore, safely neglected in our modelling.} A multi-wavelength spectrum of {the core emission from \src} extending from GHz radio frequencies up to TeV $\gamma$-rays was compiled using the data described in the previous section. To explain the SED we invoke: i) a compact (sub-pc scale) region for the non-thermal emission from IR wavelengths to VHE $\gamma$ rays, ii) an extended (pc-scale) region for the radio ($\sim10$--100~GHz)  emission, and  iii) the host galaxy of \src for the optical/UV thermal emission.

\subsection{High-energy emitting region}
We assume that the region responsible for the core emission of \src (from IR wavelengths up to VHE $\gamma$-rays) can be described 
as a spherical blob of radius $r^\prime_{\rm b}$, containing a tangled magnetic field of strength $B^\prime_{\rm b}$ and moving towards us with a Doppler factor $\delta_{\rm D}$\footnote{Henceforth, quantities measured in the rest frame of the blob are denoted with a prime.}. Protons and (primary) electrons are assumed to be accelerated to relativistic energies and to be subsequently injected isotropically in the volume of the blob at a constant rate. The latter translates to a particle injection luminosity $L^{\prime}_{\rm i}$  that can be written in dimensionless form as $\ell_{\rm i} = \sth L^{\prime}_{\rm i}/4\pi r^\prime_{\rm b} m_{\rm i}c^3$, where $i=e,p$. The accelerated particle distributions at injection are modelled generally as broken power laws, namely 
$N^\prime_{\rm i}(\gamma^\prime)\propto A_{\rm i,1}\gamma^{\prime -p_{\rm i,1}}$ for $\gamma^\prime_{\min,i} \le \gamma^\prime < \gamma^\prime_{\rm br, i}$ and $N^\prime_{\rm i}(\gamma^\prime)\propto A_{\rm i,2}\gamma^{\prime -p_{\rm i,2}} {\rm e}^{-\left(\gamma^\prime/\gamma^\prime_{\rm i, \max}\right)^b}$ for 
$\gamma^\prime_{\rm br, i} < \gamma^\prime < \gamma^\prime_{\max,i}$, where $b$ is the steepness of the exponential cutoff \citep[e.g.][]{lefa11} and $A_{\rm i, {1,2}}$ are normalization constants.  

The production of mesons, namely pions ($\pi^{\pm}, \pi^0$), muons ($\mu^{\pm}$) and kaons ($K^{\pm}, K^0$), is a natural outcome 
of $p\pi$ interactions taking place between the relativistic protons and the internal photons; the latter are predominantly synchrotron photons
emitted by relativistic electrons\footnote{Henceforth, we refer to electrons and positrons commonly as electrons.}. At any time, the relativistic electron population is comprised of those that have undergone acceleration ({\sl primary}) and those that have been produced by other processes ({\sl secondary}).
These include (i) the decay of $\pi^{\pm}$, i.e. $\pi^{+}\rightarrow \mu^{+}+\nu_{\mu}$, $\mu^{+}\rightarrow e^{+}+\bar{\nu}_{\mu}+\nu_{\rm e}$, (ii) the direct production through the Bethe-Heitler process $p \gamma \rightarrow p + e^{-} + e^{+}$ and (iii) the photon-photon absorption ($\gamma \gamma$) $\gamma \gamma \rightarrow e^{+}+e^{-}$. In addition, $\pi^0$ decay into VHE $\gamma$-rays (e.g. $E_{\gamma}\sim10$~PeV, for a parent proton with energy $E_{\rm p}=100$~PeV), and those are, in turn, susceptible to ($\gamma \gamma$) absorption and can 
initiate an electromagnetic cascade \citep{mannheim91}. {\sl It is the synchrotron radiation of secondary electrons that has a key role in shaping the high-energy part of the SED in \src } (see Sec.~\ref{sec:results}). 

Relativistic neutrons and neutrinos ($\nu_{\mu}, \nu_{\rm e}$) are also produced in $p\pi$ interactions and together with photons, electrons and protons complete the set of the five stable particle populations, that are at work in the compact emitting region of the blazar. 
We note that all particles are assumed to escape from the emitting region in a characteristic timescale, which is set equal to the photon crossing time of the source, i.e. $t^\prime_{\rm i, esc}=r^\prime_{\rm b}/c$. This energy-independent term mimics the adiabatic expansion of the source, since
the steady-state particle distributions derived by solving a kinetic equation containing a physical escape term or an adiabatic loss term are similar. 

The interplay of the processes governing the evolution 
of the energy distributions of the five stable particle populations is 
formulated with a set of five time-dependent, energy-conserving kinetic equations. 
To simultaneously solve the coupled kinetic equations for all 
particle types we use the time-dependent code described in \citet{DMPR2012}. 

\subsection{Radio emitting region}
As we show in Sec.~\ref{sec:results} the emission from  the compact emitting region of \src  cannot explain the Planck and MOJAVE radio observations, as the synchrotron spectrum is self-absorbed at $\nu \lesssim 10^{12}$~Hz; this is a common feature of single-zone models for blazar emission (see e.g. Figs. 7-10 in \citet{celotti_ghisellini08} for leptonic models and \citet{boettcher_reimer13, petrodimi15} for hadronic models). Radio emission at GHz frequencies can  be more naturally explained as synchrotron radiation produced from a larger and less compact region, such as the base of the pc-scale jet, i.e. the radio core \citep[][]{lister13}. The pc-scale jet can be modeled using a series of a (large $\sim 50$) number of  slices, where the magnetic field strength and particle distributions are assumed to evolve self-similarly along the jet  \citep[see e.g.][]{hervet_boisson15}. 
As our focus is the broad HE spectrum of \src, we do not adopt a sophisticated jet model for the radio emission \citep[e.g.][]{potter_cotter12,potter_cotter13}. Instead, we assume that the radio emission is produced by a spherical region of pc-scale size, further away from the central engine.  Our choice and a possible physical interpretation are discussed in Sec.~\ref{sec:discussion}. 
We have also verified that the external Compton scattering of the synchrotron radiation field from the extended radio blob by the electrons in the sub-pc blob can be neglected for the adopted parameters (see Tables~\ref{tab-0}-\ref{tab-1} and Appendix in \citet{petro14}). On the other way around, the radiation produced by the sub-pc blob is negligible for EC by the electrons in the pc-scale region, as long as their separation distance is $\gtrsim 0.3$~pc.

\subsection{Host galaxy radiation}
The host galaxy of \src has been clearly detected in the (near-infrared) H-band \citep{kotilainen_98} and in the (blue) B-band \citep{hyvonen_07}. 
The multi-colour imaging results suggest that the host galaxy of \src, as for most low-redshift BL Lac {objects}, is a luminous and massive elliptical galaxy; its mass has been estimated to be $10^{11.4\pm0.03} M_{\odot}$ \citep{woo05}. 
\citet{hyvonen_07} showed that the luminosity ratio between the nuclear region and the host galaxy is 1.7 and 0.58 in the U and B bands. Since 
the contribution of the host galaxy cannot be neglected we will model it based on an appropriate galaxy template. 
In particular, we use the template
of an elliptical galaxy produced by a single episode of star formation with age 13 Gyr, total mass $6\times 10^{11} M_{\odot}$, and solar metallicity \citep{silva_98}. The normalization of the spectrum has been ajusted to fit the data. In particular, the fluxes of the template were multiplied by a factor of $2.5/6$ to account for the mass difference between the real and simulated host galaxy.
\subsection{Accretion disk and BLR radiation}
The accretion disk can illuminate the gas in close proximity of the black hole, thus giving rise to the emission from the BLR. These two external radiation fields can serve as target photons for inverse Compton scattering by the electrons of the sub-pc scale region. Different analyses of the faint H$\alpha$ emission lines result in different estimates of the BLR luminosity, ranging between $L_{\rm BLR}=2.2\times10^{41}$ erg s$^{-1}$ \citep{stickel93} and $9.5\times10^{41}$~erg s$^{-1}$ \citep{morris88}. Nevertheless, both estimates suggest a low-luminosity BLR. Assuming a BLR covering factor of $\xi=1\%$ as inferred for other BL Lac {objects} \citep{stocke11, fang14}, the accretion disk luminosity is given by $L_{\rm disk}\lesssim 10^{44}\, L_{\rm BLR, 42}\, \xi^{-1}_{-2}$~erg s$^{-1}$. The inner radius of the BLR can be also estimated as $R_{\rm BLR}\simeq 3\times 10^{16}\, L_{\rm disk,44}^{1/2}$~cm \citep{ghisellini_tavecchio08}. Being conservative, we do not include in our modelling any external 
photon fields. In 
fact, as we detail in Sec.~\ref{sec:results} the HE and VHE $\gamma$-ray emission of \src  can be explained alone with radiation produced internally. 
Inclusion of the accretion disk and BLR radiation fields in the fitting procedure might result in different parameter values but it would not alter our main conclusions. 
\begin{table}
\centering
\caption{Model parameters describing the sub-pc scale region producing the non-thermal IR-TeV emission of \src. {The parameters describing
the particle distributions refer to those at injection.}}
\begin{threeparttable}
\begin{tabular}{ccc}
\hline
Parameters & model A   & model B  \\
\hline
$B^\prime$ (G) &   2.5   &  5 \\
$r_{\rm b}^\prime$ (cm)     &   $10^{15}$ & $3\times 10^{15}$ \\
$\delta_{\rm D}$            &   12.2 & 13.5 \\
$\theta$(\textdegree)\tnote{\textdagger}  & 3.8  & 3.2 \\
$\Gamma$\tnote{\textdagger}               &   7.7  & 8.2 \\
$\ell_{\rm e}$        &   $4\times10^{-4}$&  $2.5\times10^{-5}$ \\
$\gamma^\prime_{\rm e, \min}$&   10 & 10   \\
$\gamma^\prime_{\rm e, br}$&     $1.2\times10^3$& $1.2\times10^3$ \\
$\gamma^\prime_{\rm e, \max}$&    $6.3\times10^3$ &  $3.2\times10^5$ \\
$p_{\rm e, 1}$&       1.6 & 1.8 \\
$p_{\rm e, 2}$&       4.0 & 4.0 \\
$b_{\rm e}$\tnote{\textdaggerdbl}   & 1.8 & 1.0 \\
$\ell_{\rm p}$        &   $4\times10^{-2}$ & $2\times10^{-3}$  \\
$\gamma^\prime_{\rm p, \min}$&   1 & 1 \\
$\gamma^\prime_{\rm p, br}$&    $-$& $-$\\
$\gamma^\prime_{\rm p, \max}$&    $1.2\times10^6$ & $5\times 10^7$\\
$p_{\rm p, 1}$&       1.6 & 2\\
$p_{\rm p, 2}$&       $-$ & $-$ \\
$b_{\rm p}$   &       1.8 & 1.0 \\
\hline
 \end{tabular}
 \tnote{\textdagger} Assuming that the apparent speed determined for the pc-scale jet, $\beta_{\rm app}=6.22 \pm 0.18c$ \citep{lister16},
 applies also to the sub-pc jet, we solve  for the bulk Lorentz factor $\Gamma$ and angle $\theta$, given the Doppler factor $\delta_{\rm D}$ derived from the SED fitting.\\
 \tnote{\textdaggerdbl} This parameter cannot be constrained; the exact value does not affect the fit, because of the steep electron distribution. We thus adopt $b_{\rm e}=b_{\rm p}$.
 \end{threeparttable}
\label{tab-0}
 \end{table}
 
\begin{table}
\centering
\caption{Model parameters of the extended (pc-scale) region used for modelling radio emission of \src.}
\begin{threeparttable}
\begin{tabular}{cc}
\hline
Parameter &  Value\\
\hline
$B^\prime$ (G) & 0.01 \\
$r^\prime$ (cm)& $1.9\times 10^{18}$\\
$\delta_{\rm D}$ &12.2\\
$\theta$(\textdegree)\tnote{\textdagger} & 3.8  \\
$\Gamma$              & 7.7 \\
$\ell_{\rm e}$        & $6.3\times10^{-7}$ \\
$\gamma^\prime_{\rm e, \min}$&  $2\times 10^2$    \\
$\gamma^\prime_{\rm e, br}$  &  $8\times10^3$  \\
$\gamma^\prime_{\rm e, \max}$&  $3\times10^4$    \\
$p_{\rm e, 1}$& 1.8\\
$p_{\rm e, 2}$& 4.0\\
$b_{\rm e}$ & $\gg 1$ \\
\hline
 \end{tabular}
 \tnote{\textdagger} Same as in Table~\ref{tab-0}.
 \end{threeparttable}
\label{tab-1}
 \end{table}
\section{Results}
\label{sec:results}
\subsection{The core emission of \src}
Aim of this paper is to demonstrate the possibility of producing broad HE non-thermal spectra from a single compact emitting region instead of deriving a unique parameter set as determined by the best $\chi^2$ fit to the data. We, therefore, present two indicative model fits to the SED of \src that differ only in the properties of the compact component.  We will refer to these as models A and B (see Table~\ref{tab-0}). The parameter values that describe the radio-emitting region are summarized in Table \ref{tab-1}. 
\begin{figure*}
\centering
\includegraphics[width=0.49\textwidth]{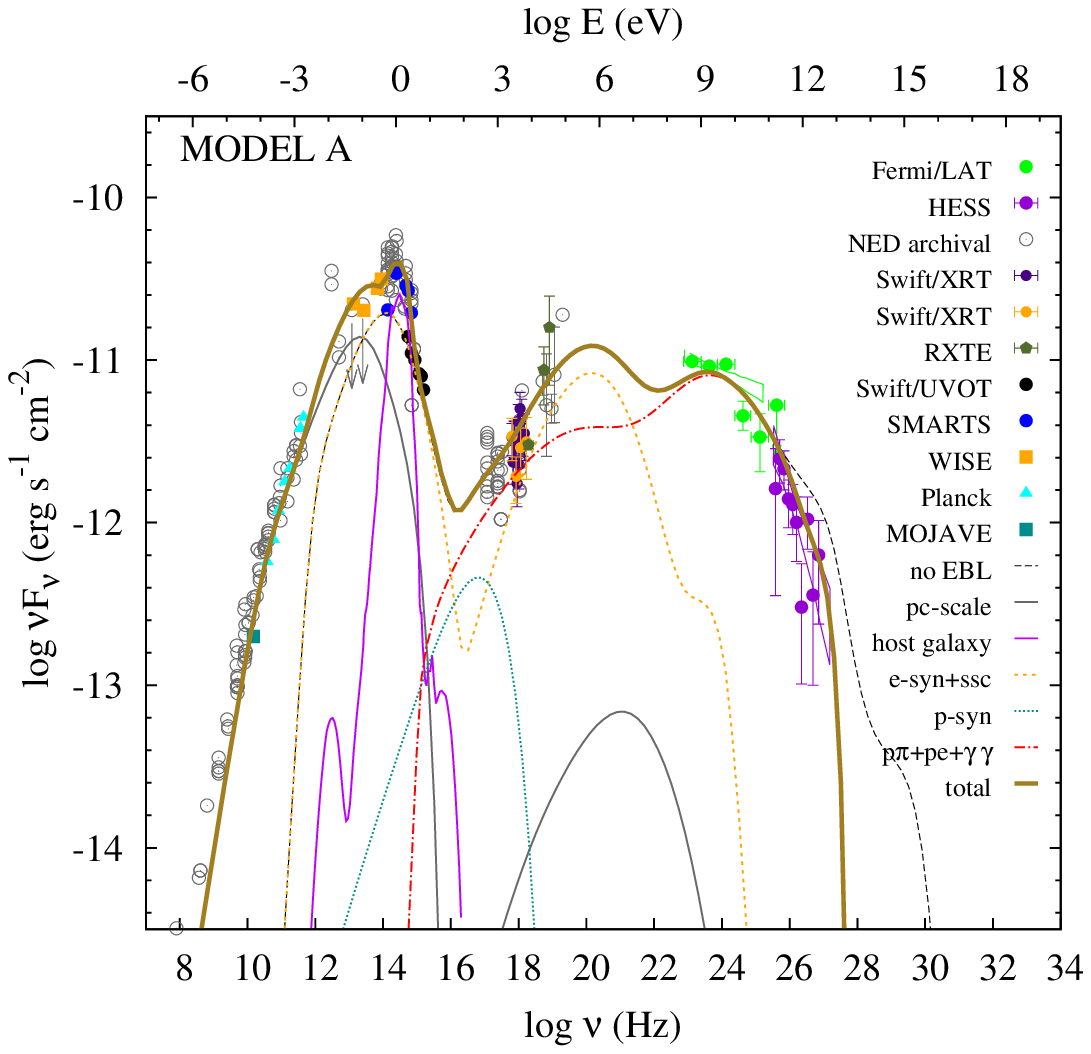} 
\includegraphics[width=0.49\textwidth]{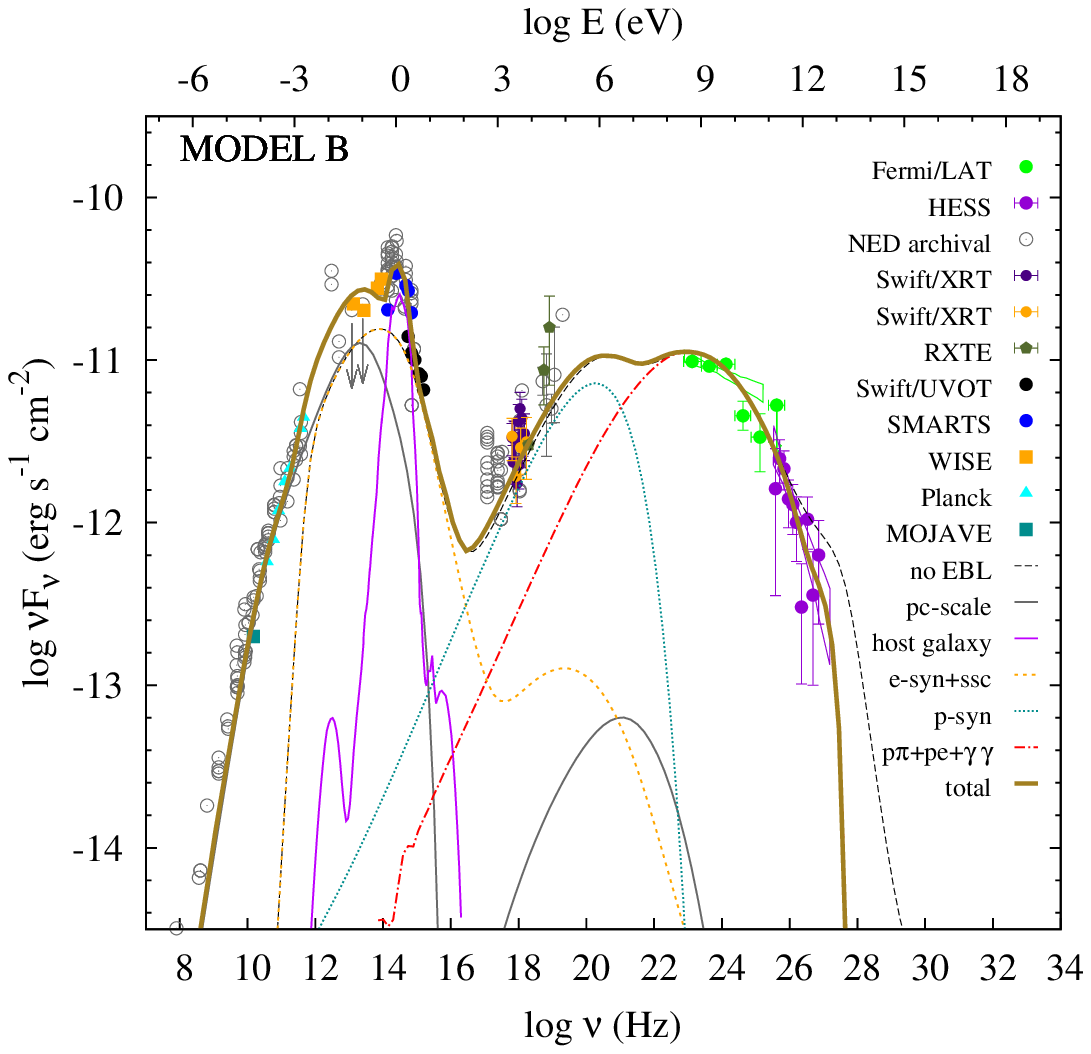} 
\caption{SED of \src \ compiled using multi-wavelength data (filled colored symbols) collected by various instruments as noted in the legend (for details, see Sec.~\ref{sec:data}). The grey open  circles depict archival data obtained from the NED database.  The multi-wavelength spectrum (thick gold line) is composed of the emission from the pc-scale radio emitting region (grey solid line),  the host galaxy (magenta solid line), and the emission from the sub-pc region. This is decomposed into the following components: SSC radiation from primary electrons (orange dashed line), proton synchrotron radiation (dark cyan dotted line), and synchrotron radiation from secondary electrons produced by $p\gamma$ and $\gamma \gamma$ processes (red  dash-dotted line).   The spectrum  without taking the EBL absorption into account is  overplotted with a black dashed line. 
 The results for models A and B are displayed in the left and right panels, respectively. {For interpretation of the references to colour in this figure legend, the reader is referred to the web version of this article.}}
\label{fig:sed1}
\end{figure*}

\begin{figure*}
\includegraphics[width=0.49\textwidth]{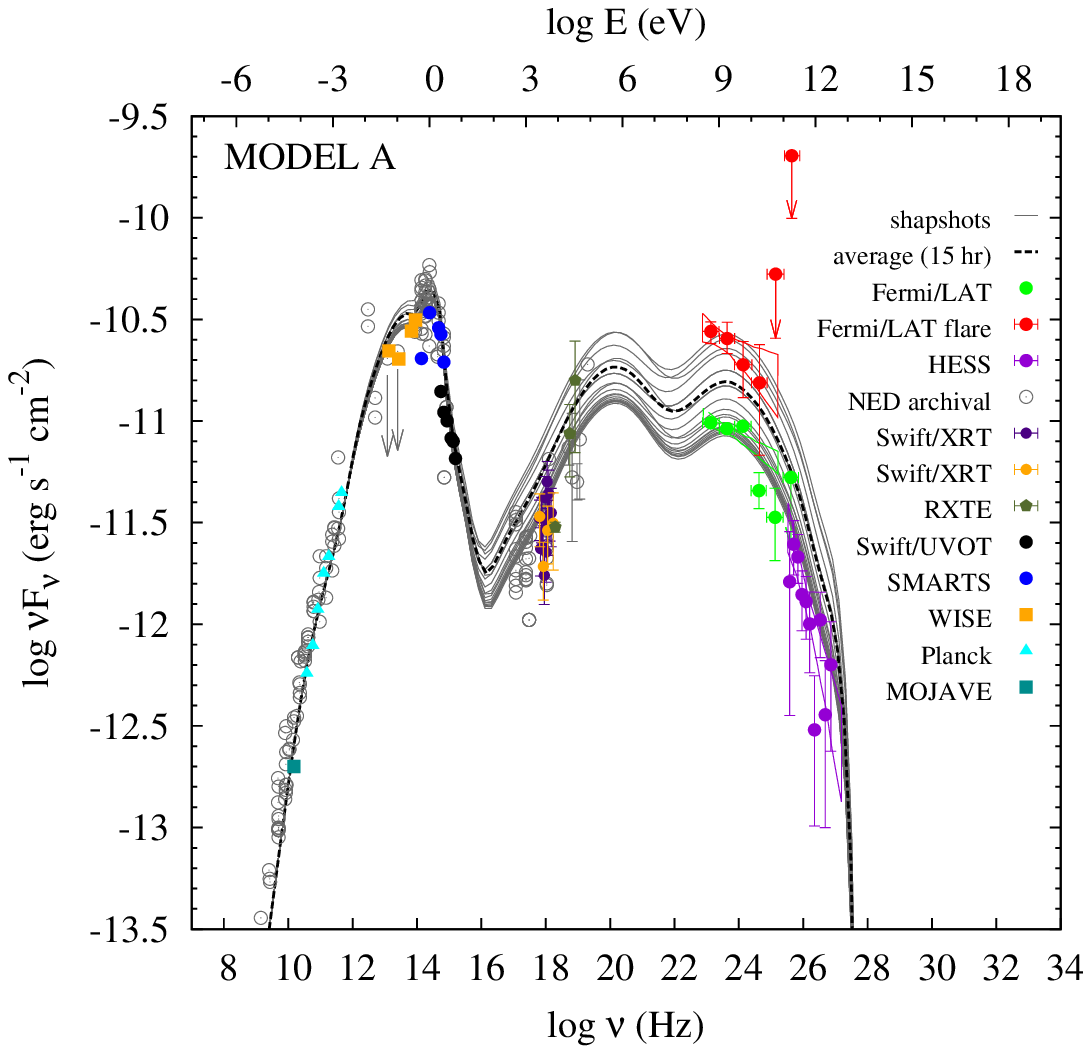}
\includegraphics[width=0.49\textwidth]{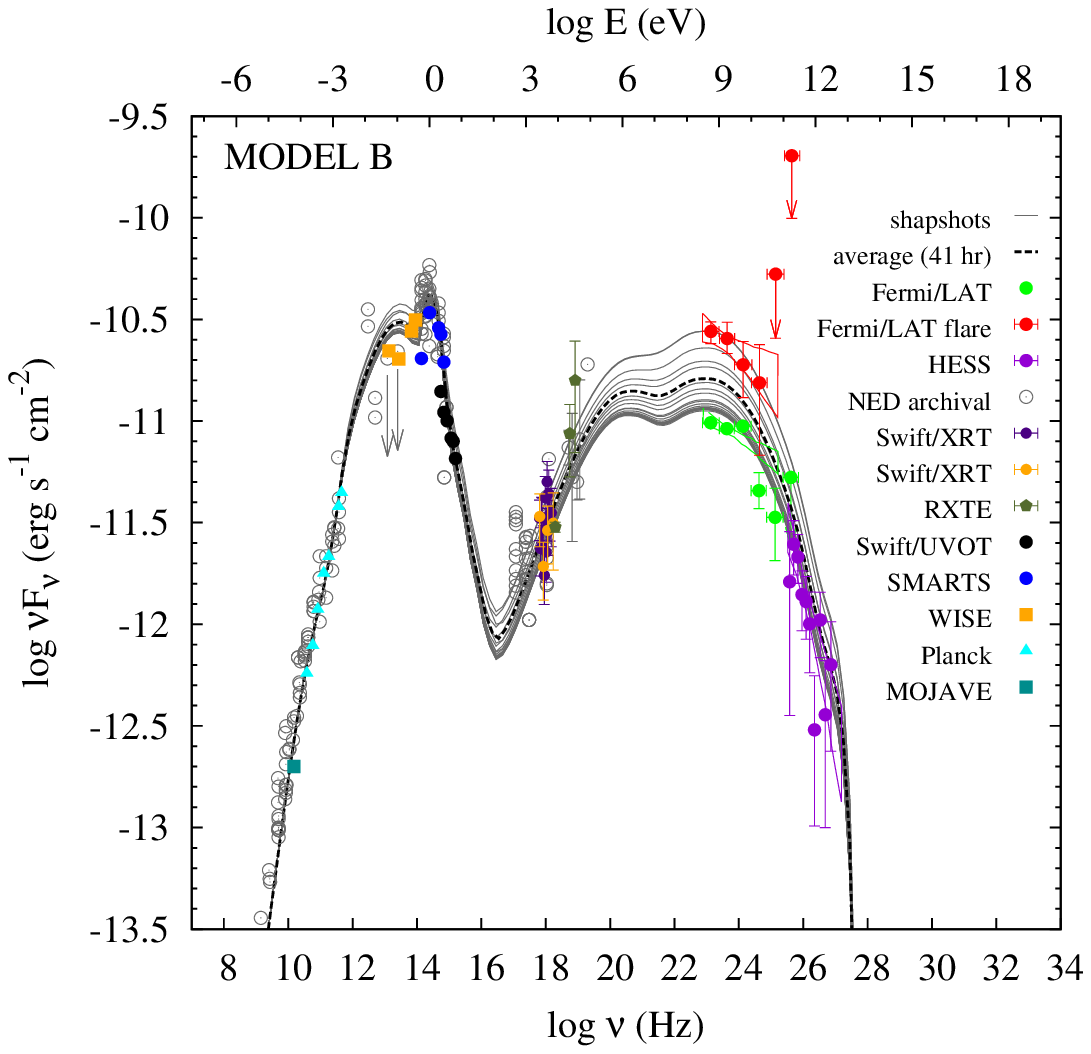}
\caption{SED snapshots (grey thin lines) as obtained in models A and B for the fiducial flare discussed in text. The time-averaged spectrum is overplotted (black dashed line). In addition to the observations shown  in Fig.~\ref{fig:sed1}, we include the time-averaged \fermi spectrum during the period MJD 56306-56376 (red symbols) for comparison reasons. {For interpretation of the references to colour in this figure legend, the reader is referred to the web version of this article.}}
\label{fig:sed_flare}
\end{figure*}

Our results for models A and B are presented, respectively, in the left and right panels of Fig.~\ref{fig:sed1}. 
The total multi-wavelength spectrum (thick gold line) is composed of the emission from the pc-scale radio emitting region (grey solid line),  the host galaxy (magenta solid line), and the emission from the sub-pc region.  The latter is decomposed into the following emission components: the SSC radiation from primary electrons (orange dashed line), the proton synchrotron radiation (dark cyan dotted line), and the synchrotron radiation from secondary electrons produced by the $p\gamma$ and $\gamma \gamma$ processes (red  dash-dotted line).  The black dashed curve shows the spectrum before the attenuation on the EBL {and indicates the degree of the internal to the source $\gamma \gamma$ absorption.} 
Both models provide a satisfactory representation of the source's SED without the need of external photon sources to account for the \fermi and \hess data. The broad and curved $\gamma$-ray spectra obtained in both models are a natural outcome of the leptohadronic scenario, despite the fact that the radiative processes responsible for the X-ray 
and $\gamma$-ray emission are different. 
This can be understood by inspection of the various emission components that comprise the total emission from the sub-pc scale region. 

The X-ray emission in model A is mainly produced by the SSC emission of primary electrons (orange dashed line) and, in this regard,
it resembles the pure leptonic SSC models (see e.g. Fig.~3 in \citet{sanchez15}). Note, however, that at $10^{17}-10^{18}$~Hz ($\sim 0.4-4$~keV) the synchrotron emission from Bethe-Heitler pairs (red dash-dotted line) is comparable to the SSC one. The synchrotron emission from secondary pairs produced by the $p\pi$ and $\gamma \gamma$ processes dominates the $\gamma$-ray emission in the \fermi and \hess energy bands, similarly to what have been shown for several other HBL \citep[e.g.][]{petrodimi15, petrocoenders16}. 
While in HBL the target photons for $p\pi$ interactions belong to the low-energy component of the SED, in the case of \src, which is an LBL, the photons of the low-energy hump are less energetic and typically cannot satisfy the energy threshold condition for pion production.
In particular,  for protons with Lorentz factors $\gamma^\prime_{\rm p}$, the energy threshold condition for pion production is satisfied by photons with observed energies $\epsilon \gtrsim  \delta_{\rm D} \bar{\epsilon}_{\rm th}/\gamma^\prime_{\rm p} \simeq 1.4 \unit{keV} \, \delta_{\rm D, 1}/\gamma^\prime_{6}$, where $\bar{\epsilon}_{\rm th}=145$~MeV. Given that the maximum energy of the protons is $\gamma^{\prime}_{\rm p, \max}\sim 10^6$ ($10^7$) in model A (model B), hard X-ray photons ($\gg$1 keV)  will serve as targets for photopion interactions with the less energetic protons. 
As the energy threshold for Bethe-Heitler pair production is lower, protons with $\gamma^\prime_{\rm p}=10^3$ ($10^6$) will interact with $\sim$keV (eV) photons. These simple estimates indicate that the resulting $\gamma$-ray emission is a  non-linear combination of various radiative processes.

In model B, because of the larger $\gamma^\prime_{\rm p, max}$ and stronger magnetic field, the proton synchrotron component dominates the soft-to-hard X-ray band (right panel in Fig.~\ref{fig:sed1}), while the synchrotron radiation  from Bethe-Heitler and $\gamma \gamma$ produced pairs contributes the most to the observed photohadronic emission (red dash-dotted line). Electrons produced by Bethe-Heitler interactions much above the threshold, attain a maximum Lorentz factor $\gamma^\prime_{\rm e, pe} \simeq 4 \gamma^{\prime 2}_{\rm p} \epsilon_{\rm t}^\prime $, where $\epsilon^\prime_{\rm t}$ is the energy of the target photon  \citep[e.g.][]{kelneraharonian08}. Thus, the characteristic synchrotron photon energy is $\propto B \gamma_{\rm p}^{\prime 2}$ (see also \citet{petromast15}). Based on the parameter values listed in Table~\ref{tab-0} and Fig.~\ref{fig:sed1} (right panel), the peak energy of the Bethe-Heitler component in model B is expected to be $\sim 3500$ larger than in model A, i.e. $\sim 0.1$~GeV.

Thus, the theoretical spectra differ significantly in the relative importance of the various emission processes, despite the small differences in the adopted parameter values of models A and B. Our results reflect the
non-linearity of the radiative processes that have to be present in a system that contains relativistic electrons and protons.
\subsection{Variable VHE core emission}
The decomposition of the SED obtained in models A and B revealed their differences in the origin of the X-ray and $\gamma$-ray emission. Different X-ray and $\gamma$-ray variability signatures are, therefore, expected. {\sl Regardless, both models predict variable VHE emission on timescales similar to those in X-rays,  in contrast to the scenarios that attribute the TeV emission to the kpc-scale jet of \src. }

To illustrate the model predictions on the variability and the broadband spectral evolution  we present an indicative example of a flaring event. 
An increase of the observed flux (i.e., flare) can be attributed, in general, to a higher injection rate of radiating  particles at the dissipation region, which, in turn, may be caused by temporal modulations of the jet power.
In the following, we model a fiducial flare caused by an increase in the
injection compactness (or, equivalently luminosity) of primary electrons and protons given by:
\eqb
\frac{\ell_{\rm i}(\tau)}{\ell_{\rm i}^{(0)}} = 1 + \frac{(A/2)^2}{(\tau-\tau_{\rm pk})^2 + (A/2)^2},
\label{eq:lum}
\eqe
where $\tau\equiv t^\prime/t^\prime_{\rm dyn} = c t^\prime/r^\prime_{\rm b}$, $\tau_{\rm pk}=5$, $A=2$ and $\ell_{\rm i}^{(0)}$ is the value derived from fitting the SED (see Table~\ref{tab-0}). Here, the starting time $\tau=0$ corresponds to the time that is needed to establish the steady-state emission shown in Fig.~\ref{fig:sed1} and $\tau_{\rm pk}$ is the peak time where $\ell_{\rm i}(\tau_{\rm pk})=2\ell_{\rm i}^{(0)}$. We define  $\Delta \tau^{\rm  inj}_{75\%} = A$, i.e. the time interval where $\ell_{\rm i} \ge 0.75 \ell_{\rm i, pk}$, to be the measure of the injection's duration.  This correspond to $\Delta t^{\rm inj}_{75\%} = A t^\prime_{\rm dyn} (1+z)/\delta_{\rm D} \simeq 1.9 \, {\rm hr} \,  r^\prime_{\rm b, 15}/\delta_{\rm D, 1}$.

\begin{figure}
\resizebox{\hsize}{!}{\includegraphics[]{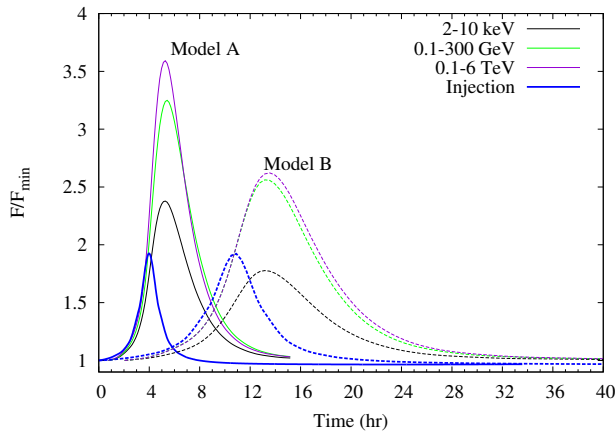}}
\caption{Model-derived light curves for the fiducial flare described in text. The results obtained in models A and B are shown with solid and dashed lines, respectively. Here, $F$ is the integrated flux at different energy bands marked on the plot and $F_{\min}$ is the respective value prior to the flare. The injection luminosity of particles (normalized to its minimum value) is also shown as a function of time (blue coloured lines).}
\label{fig:lc}
\end{figure}
Fig.~\ref{fig:sed_flare} shows, in total, 20 snapshots (grey thin lines) of the broadband emission during the flaring episode described above. Left and right panels correspond to models A and B, respectively.  The time-averaged spectrum is also plotted (black dashed line), while no attempt in fitting the \fermi flare (red points) has been made. We find that changes in the luminosity of radiating particles alone do not affect the spectral shape either in the X-ray or the $\gamma$-ray bands. Interestingly, no significant spectral change was detected during the \fermi flare of 2013 (MJD 56306-56376) \citep[][]{abramowski15}. The flux variations in the optical/UV energy bands are less pronounced than these at higher energies, although the radiating primary electrons are more energetic than those emitting at $\sim0.1-1$~keV X-rays. The reason is that the observed optical/UV emission has a significant contribution from the host galaxy itself, while the primary synchrotron component is sub-dominant 
(see Fig.~\ref{fig:sed1}).

The integrated flux in different energy bands, normalized to its pre-flare value ($F_{\min}$), is presented as a function of time in Fig.~\ref{fig:lc}. 
For comparison reasons, the ratio $\ell_{\rm i}/\ell_{\rm i}^{(0)}$ is also shown (blue coloured lines). In both models, we find no time-lag between the X-ray and $\gamma$-ray energy bands.  Unless there is a time-lag in the injection of accelerated electrons and protons, the flares predicted by the models are (quasi)-simultaneous. The properties of the light curves obtained in models A and B, namely peak flux $F_{\rm pk}$ and duration  are summarized in Table~\ref{tab-2}. The amplitude of the flare is larger at higher energies and its shape becomes more symmetric, i.e. the rise and decay timescales are similar. In both models, the rise timescale of the flares in X-rays and $\gamma$-rays are similar. On the contrary, the $\gamma$-ray flares appear to be  shorter in duration compared to those in X-rays.  {In particular, we find that the X-ray flare is twice as long as the injection episode, namely $\Delta t_{75\%}\sim 4 t_{\rm dyn}$, where $t_{\rm dyn}\simeq 0.75$ hr (2.11 hr) is the crossing time of the 
source in the observer's frame for Model A (Model B). The $\gamma$-ray flare is shorter than the X-ray flare by one $t_{\rm dyn}$ in both models. Because of the superposition of various emitting components at different energy bands, it is not straigthforward to provide an explicit expression for the expected flare duration. This can be, however, qualitatively understood; } the faster decay timescale of the $\gamma$-ray flares reflects the shorter cooling timescale of the radiating (secondary) electrons, which are typically more energetic than those emitting in X-rays \citep[see Section 3.1 in][]{petromast15}.

Although the injection luminosity function for primary electrons and protons is the same in models A and B, the obtained peak fluxes are lower in the latter. The differences are related to the underlying physical process responsible for the X-ray and $\gamma$-ray emission.
For example,  the X-ray emission in model B is expected to be $\propto \ell_{\rm p}$, since it is dominated by the proton synchrotron radiation. On the contrary, the X-ray variability  amplitude in model A is expected to be larger than in model B, since the X-ray emission is a superposition of the SSC and Bethe-Heitler components that respectively dependent on the varying $\ell_{\rm e}$ and $\ell_{\rm p}$.

{An interesting point to be considered is the detectability of the X-ray flux variability with \swift/XRT during a fiducial $\gamma$-ray flare, as shown in Fig.~\ref{fig:lc}. The source has been so far observed with \swift/XRT for a few ks, with single observations being usually split to several (two to four) snapshots with exposure times less than 1 ks. Based on the count rate of the existing \swift/XRT observations, a one ks exposure would deliver approximately 100 counts. A flux increase by a factor of two, as shown in Fig.~\ref{fig:lc}, would be detectable with \swift/XRT above a $3\sigma$ 
level. This could be achieved with multiple observations (of at least one ks exposure time) spanning over the duration of the fiducial flare. The detectability of a TeV flare with the next generation of IACTs is discussed in the following section. }
 \begin{table}
\centering
\caption{Amplitude and duration of the fiducial flare described in text at X-rays (2-10 keV), HE (0.1-300 GeV) and VHE (0.1-6 TeV) $\gamma$ rays. The respective fluxes prior to the flare ($F_{\min}$) are also listed.}
\begin{tabular}{cccc}
\hline
    &  $2-10\,{\rm keV}$ & $0.1- 300 \,{\rm GeV}$ & $0.1- 6 \,{\rm TeV}$\\           
\hline
Pre-flare flux:& \multicolumn{3}{c}{$F_{\min}$ (erg cm$^{-2}$ s$^{-1}$)} \\
model A & $6.8\times 10^{-12}$ & $4.8\times 10^{-11}$ & $5.2\times 10^{-12}$ \\ 
model B &  $3.7\times 10^{-12}$ & $5.5\times 10^{-11}$ & $4.1\times 10^{-12}$ \\ \\
Amplitude: & \multicolumn{3}{c}{$F_{\rm pk}/F_{\min}$} \\
model A& 2.4 & 3.2 & 3.6\\
model B& 1.8 & 2.5 &2.6\\  \\
Duration: & \multicolumn{3}{c}{$\Delta t_{75\%}$ (hr)} \\
 model A& 3.0 & 2.3 & 2.3\\
 model B& 8.6 & 6.6& 6.5\\
\hline
 \end{tabular}
\label{tab-2}
 \end{table}
\begin{figure*}
\centering
2\includegraphics[width=0.49\textwidth]{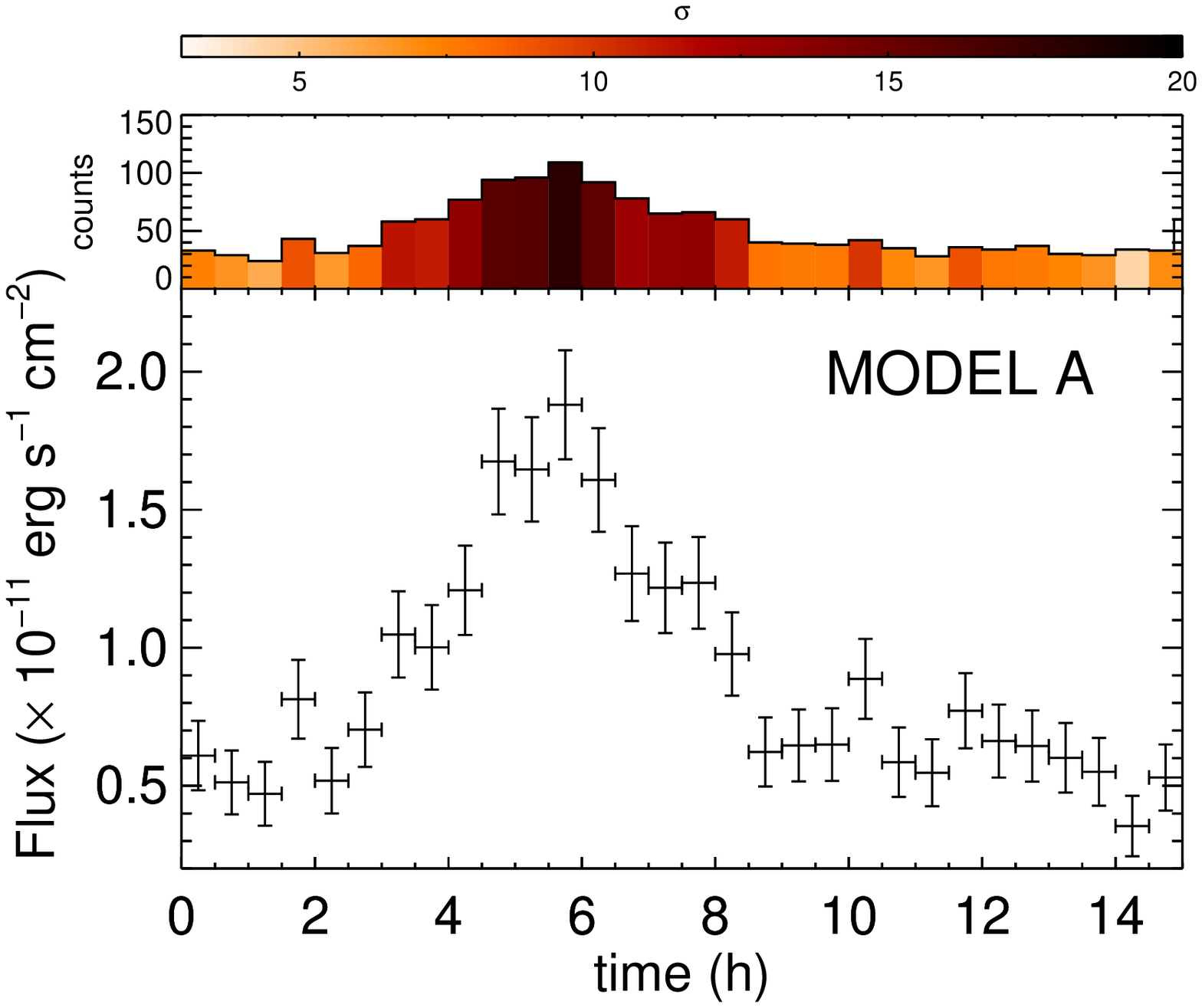} 
\includegraphics[width=0.49\textwidth]{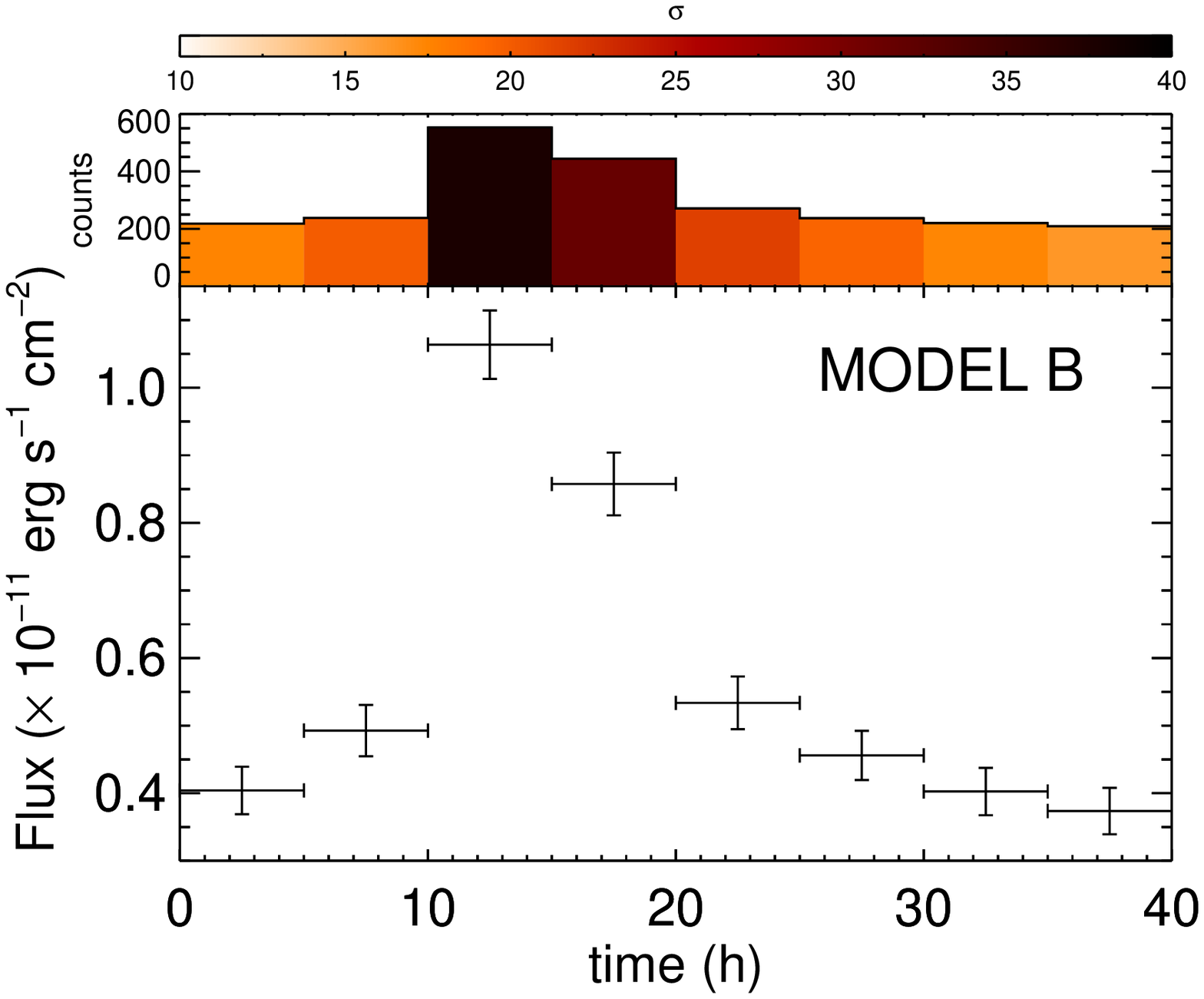} 
\caption{Simulated CTA light curves for models A (left panel) and  B (right panel). The event files were generated according to the 0.1-6 TeV light curves shown in Fig.~\ref{fig:lc}. A time binning of 0.5 h and 5 h was used for the simulated light curves of models A and B, respectively. The expected number of source counts is shown in the top panel of both plots. Colour coding is used for the significance  of the detection (in Gaussian $\sigma$) as obtained by the analysis of the simulated event files.}
\label{fig:cta_LC}
\end{figure*}

\section{Prospects for CTA}
\label{sec:cta}
\src has been one of the few VHE emitting LBL objects and it was first discovered by \hess \citep{hofmann10,abramowski15}. As discussed by the authors, \hess was able to detect the system with a significance of 6.6 $\sigma$ during an integrated exposure time of 14 h. The Cherenkov Telescope Array \citep[CTA,][]{actis_11, 2013APh....43....3A} is a next-generation observatory of Imaging Air Cherenkov Telescopes (IACT). {It is planned to cover} more  than $1$~km$^2$ area and will be composed of an array of large, middle, and small-sized telescopes. When completed, CTA {is expected to} reach an effective area larger than the Cherenkov light pool size, and deliver a sensitivity about an order of magnitude better than that of current Cherenkov telescopes. In principle, CTA will be therefore able to detect \src at a fraction of the time needed by \hess

In order to test this we created simulated light curves of the VHE  flares (0.1-6 TeV) predicted by the models A and B (Fig.~\ref{fig:lc}) using the software package \textsc{ctools}\footnote{http://cta.irap.omp.eu/ctools/}  \citep{GammaLib16}. Our simulated events are drawn from three components i) a point source with the spectral properties of \src, ii) an isotropic CR background that was modeled as a diffuse isotropic source with a spectral shape and flux  adopted by \cite{silverwood_15} (see Fig.~2 therein), and iii) an instrumental background of the detector (see \textsc{CTAIrfBackground} of \textsc{ctools}).
We used the task \textsc{ctobssim} to simulate the event files, and the \textsc{ctlike} tool to perform a maximum likelihood fitting of a power-law model (photon index 2.65) to the unbinned simulated data. Finally the \textsc{cttsmap} tool was used to confirm the significance of the detection. The procedure was repeated for various flux levels of the models shown {in} Fig. \ref{fig:lc}. The simulated data were binned in intervals of 0.5 h for model A and 5 h for model B. In both cases, the adopted bin size is less than the maximum visibility of the system during a single day. 

The results of our simulations are presented in Fig.~\ref{fig:cta_LC}. For a TeV flare with $\sim 2.3$~hr duration and flux increase by factor of two compared to the quiescent flux level (model A; see Table~\ref{tab-2}), we find that such variability would be detected by CTA even with short exposure times (0.5 hr). In particular, at the peak time of the flare the significance of the detection would exceed $15\, \sigma$ (see left panel in Fig.\ref{fig:cta_LC}). A detection significance similar to that of \hess ($\sim 6\, \sigma$)  would be achieved for the quiescent flux levels  but at a fraction of the integrated exposure time of \hess A longer exposure time (5 hr) was adopted for the longer duration flare predicted by model B. Despite the lower peak flux of the flare (see also Table~\ref{tab-2}), the expected number of counts at the peak time of the flare is six times larger than that of flare A due to the larger exposure time. For a 5~hr exposure time, the lowest significance that can be reached is still 
above 
$15\,\sigma$. 

\section{Discussion} 
\label{sec:discussion}
We have presented two indicative models for explaining the broad high-energy spectrum of \src with their parameters listed in Table~\ref{tab-0}. Both models are viable alternatives, when only considering their ability of reproducing the observed SED. 
However, they differ in terms of energetic requirements. 
The power of a two-sided jet can be written as \citep[e.g.][]{ghisellini14} 
$P_{\rm j} =  2\pi r_{\rm b}^{\prime 2} \Gamma^2 c \sum_{\rm i}u^\prime_{\rm i} + (8/3)(\Gamma^2/\delta_{\rm D}^4)L_{\rm ph}$, where $u^\prime_{\rm i}$ ($i=e,p,B$) is the energy density as measured in the respective rest frame, $L_{\rm ph}$ is the apparent photon luminosity and the last term in the right hand side of the equation is the bolometric absolute photon power \citep[e.g.][]{dermer_murase12}. The inferred jet power for models A and B is respectively $10^{48}$ erg~s$^{-1}$ and $10^{47}$~erg s$^{-1}$ to be compared to the jet radiation power $P_{\rm r}\sim 3\times10^{42}$ erg~s$^{-1}$ and the Eddington luminosity of \src $L_{\rm Edd}\sim 3\times 10^{46}$~erg s$^{-1}$, for a black hole mass $M_{\rm BH}=10^{8.4\pm 0.06} M_\odot$ \citep{woo05}. 
It is interesting to note that even in scenarios that invoke the presence of relativistic electrons alone, the  jet power may be as high as $\sim 10^{47}$~erg s$^{-1}$ \citep[see e.g.][]{hervet_boisson15}. 

If the accretion operates in the magnetically arrested (MAD) regime \citep{narayan03}, the jet power may be related to the accretion power, $\dot{M}c^2$,  as $P_{\rm j} = \eta_{\rm j} \dot{M} c^2$, where $\eta_{\rm j} \sim 1.5-3$ \citep{sasha_12, mckinney_12} with higher $\eta_{\rm j}$ values obtained for faster spinning black holes and thicker disks. Adopting $\eta_{\rm j}=2$, the accretion power for models A and B is, respectively, $5\times10^{47}$ erg~s$^{-1}$ and $5\times 10^{46}$ erg~s$^{-1}$. The luminosity of the accretion flow can be estimated from the BLR luminosity assuming a covering fraction $\xi$, i.e., $L_{\rm disk}\lesssim 10^{44}\, L_{\rm BLR, 42}\, \xi^{-1}_{-2}$~erg s$^{-1}$ and the radiative efficiency is given by 
$\epsilon=L_{\rm disk}/\dot{M} c^2=2\times 10^{-4}$ ($2 \times 10^{-3}$) for model A (model B). The low radiative efficiency in this source is not a feature unique to our model \citep[see also][]{hervet_boisson15}. 

Assuming that the magnetic field is mostly toroidal at pc scales it can be written as $B^\prime=(4 P_{\rm B}/c)^{1/2}(R\Gamma\theta_{\rm j})^{-1}$, where  $P_{\rm B}$ is the jet power carried by the magnetic field, $R \simeq r^\prime_{\rm b}/\theta_{\rm j}$ is the distance from the black hole, and $\theta_{\rm j}$ is the opening angle of the jet. 
At sub-pc scales we found that most of the contribution to the jet power comes from the relativistic proton component, i.e., $P_{\rm j}\simeq P_{\rm p}$. Assuming that the ratio $P_{\rm B}/P_{\rm p}$ remains constant from the sub-pc to the pc scales and equal to $\sim 10^{-3}$ ($10^{-6}$) for model B  (model A), the magnetic field at $R=5$ pc is estimated to be $16$~mG (3 mG) for $\Gamma \theta_{\rm j}\sim \Gamma \theta \sim 0.5$ (see Table~\ref{tab-0}). Thus, the magnetic field strength in both models is comparable to the adopted value for the pc-scale emitting region in this work (see Table~\ref{tab-1}). Alternatively, it could be that $P_{\rm B}\lesssim P_{\rm j}$ at the pc scales of the jet. However, the estimated magnetic field would be then close to 1 G, i.e. larger than the adopted value at Table~\ref{tab-1} (see also \citet{pushkarev_12}). 

In general, the energetic comparison of the two models suggests that solutions with higher magnetic field strengths in the (sub-pc) emitting region are favoured. Here, we did not aim at finding the most ``economic model'' that describes the SED of \src, so we cannot exclude models with sub-Eddington accretion and jet powers \citep[see e.g.][]{petro_dermer16}. Other solutions characterized by lower jet powers, higher magnetic field strengths and larger blobs are also expected to be closer to equipartition ($P_{\rm B}\sim P_{\rm p}$). We plan to search for the model that minimizes the jet power by scanning the available parameter space in the future. In summary, our results outline a physical picture where the jet is initially Poynting-flux dominated, it dissipates a significant fraction of its magnetic energy to relativistic protons at sub-pc scale distances, and extends to pc scale distances with a constant ratio of magnetic-to-(relativistic) particle powers.

 The low-energy spectrum from the compact, high-energy emitting blob cuts off at $\sim 10^{12}$~GHz, being unable to explain the radio observations at lower frequencies. To account for the radio emission in \src, we assumed that this originates from a more extended region of the jet, which we approximated by a spherical blob with characteristic radius  $r^\prime\sim 1$~pc; a more detailed description of the pc-scale jet lies out the scope of the present paper. It is intriguing to discuss the possibility of a physical connection between the sub-pc and pc-scale blobs. On the one hand, the electron and magnetic field energy densities in the pc-scale blob are $\sim 10^5-10^6$ times smaller than those inferred for the compact region (see Table~\ref{tab-2}). Since $r^\prime/r^\prime_{\rm b}\sim 10^3$, this is compatible with a scenario of a conical jet where the energy densities are expected to decrease as $\sim 1/r^{\prime 2}$.
 On the other hand, without new injection of electrons, it is difficult for a blob that produces a high-energy flare (in X-rays and $\gamma$-rays) to also produce a radio flare later, after it has sufficiently expanded. The reason is that the electrons that are responsible for the radio emission are typically very energetic and not the result of excessive adiabatic cooling.  However, if the electron injection continues from the sub-pc to the pc-scale jet, a delayed radio flare with respect to the $\gamma$-ray one may be expected on a timescale of $3\times10^6  \, {\rm s} \, r^\prime_{18}/\delta_{\rm D,1}$ (see Table~\ref{tab-1}). In this scenario, therefore, fast ($\sim$hr) X-ray and $\gamma$-ray flares caused by a strong jet episode may be followed by radio flares a few months later \citep[see][for Mrk 421]{hovatta_15}.

\section{Summary}
\label{sec:summary}
 We have shown that the superposition of  different emission components related to photohadronic interactions can explain the HE and VHE $\gamma$-ray emission  of \src without invoking external radiation fields. This was exemplified with two indicative model fits to the SED of \src where the VHE emission was assumed to originate from the core of the jet, i.e. from a compact, sub-pc scale region. 
 Our model for the non-thermal emission of \src predicts (quasi)-simultaneous flares at X-rays, HE, and VHE $\gamma$-rays. The flare duration in the aforementioned energy bands is of the same order of magnitude, with shorter durations and larger variability amplitudes  obtained at higher energies. In addition,  no spectral changes during the flares are expected, unless the slope of the radiating particles changes. We showed that CTA would be able to detect $\sim$hr timescale variability at $E_{\gamma}>0.1$~TeV at high significance with shorter exposure times than current Cherenkov telescopes. Detection of flux variability in {GeV $\gamma$-rays and/or X-rays} could be therefore used to trigger pointing observations of \src with CTA. The detection of VHE variability on similar timescales as those observed in X-rays and GeV $\gamma$-rays would point towards a common emitting region of sub-pc scale. Although it could not rule out a kpc-jet origin of the quiescent VHE emission, as the latter could still be 
explained by an additional 
emitting component, a model of a sub-pc scale origin would be preferred in the spirit of Ockham's razor.

\section*{Acknowledgments}
We thank Dr. Tullia Sbarrato for useful discussions. 
M. P. acknowledges support from NASA through the Einstein Postdoctoral 
Fellowship grant number PF3~140113 awarded by the Chandra X-ray 
Center, which is operated by the Smithsonian Astrophysical Observatory
for NASA under contract NAS8-03060.
G.\,V. acknowledges support from the BMWi/DLR grants FKZ 50 OR 1208.
D.\, G. acknowledges
support from NASA through grant NNX16AB32G issued through the Astrophysics Theory Program.
This research has made use of the RXTE/PCA python script pca.py developed by J.-C. Leyder and J. Wilms, freely available from the HEAVENS webpage.
\bibliographystyle{mn2e} 
\bibliography{aplib}

\end{document}